\documentstyle[prl,aps,multicol,graphicx]{revtex}

\begin{document}
%\draft
%\documentstyle[preprint,aps]{revtex}
%\documentstyle[floats,prl,aps,twocolumn,graphicx]{revtex} 
%\draft

\title{Superconductivity in Ropes of Single-Walled Carbon Nanotubes}
\author{M. Kociak$^1$, A.Yu. Kasumov$^{1,2,3}$, S. Gu\'eron$^1$, B. Reulet$^1$, 
I.I. Khodos$^3$,\\ Yu.B. Gorbatov$^3$,
V.T. Volkov$^3$, L. Vaccarini $^4$ and H. Bouchiat$^1$}

\address{$^1$Laboratoire de Physique des Solides, Associ\'e au CNRS,
B\^at. 510, Universit\'e Paris--Sud, 91405, Orsay, France.$^2$ Present address: Starlab, Brussels, Belgium
$^3$Institute of Microelectronics Technology and High Purity Materials,
Russian Academy of Sciences, Chernogolovka 142432 Moscow Region,
Russia.$^4$ Groupe de Dynamique des Phases Condens\'ees, Universit\'e
Montpellier II 34095 Montpellier France.\\}
\maketitle

%\\%
%\parbox{14cm}{\medskip\rm\small%
\begin{abstract}
We report measurements on ropes of Single Walled Carbon Nanotubes (SWNT) in low-resistance contact to non-superconducting (normal) metallic pads, at low voltage and at temperatures down to 70 mK. In one sample, we find a two order of magnitude resistance drop below 0.55 K, which is destroyed by a magnetic field of the order of 1T, or by a d.c. current greater than 2.5 $\mu A$. These features strongly suggest the existence of superconductivity in ropes of SWNT.\\
%Pacs 74.10.+v, 74.70.-b, 74.50.+r
\end{abstract}


%}}
%\begin{document}

%\maketitle 
%\newpage
\begin{multicols} {2}
\narrowtext

Metallic carbon nanotubes are known to be model systems for the study of 1D electronic transport \cite{Dresselhaus,Hamada,Wildoer}. Electronic correlations are expected to lead to a breakdown of the Fermi liquid state. Nanotubes should then be described by Luttinger Liquids (LL) theories \cite{Egger,Kane}, with collective low energy excitations and no long range order.  
Proof of the validity of LL description in ropes was given by the measurement of a resistance diverging as a power law with temperature down to 10~K \cite{Bockrath}. However, this measurement was done on nanotubes separated from measuring leads by tunnel junctions. Because of Coulomb blockade \cite{Grabert}, the low temperature and voltage regime were not explored. 
In contrast, we have developed a technique in which measuring pads are connected through low contact resistance to suspended nanotubes \cite{Kasumov2}. We previously showed that when the contact pads are superconducting, a large supercurrent can flow through nanotubes \cite{Kasumov}. In this letter, we report experimental evidence of intrinsinc superconductivity below 0.55~K in ropes of carbon nanotubes connected to normal contacts.

The samples are ropes of SWNT suspended between normal metal contacts (Pt/Au bilayers). The SWNT are prepared by an electrical arc method with a mixture of nickel and yttrium as a catalyst \cite{Journet,Vaccarini}. SWNT with diameters of the order of 1.4 nm are obtained. They are purified by the cross-flow filtration method \cite{Vaccarini}. The tubes are usually assembled in ropes of a few hundred parallel tubes. Isolation of an individual rope and connection to measuring pads are performed according to the procedure we previously used \cite{Kasumov2}, where ropes are soldered to melted contacts. The contact resistance is low and the tubes can be structurally characterized with a transmission electron microscope (TEM). For the three samples presented here, the contacts were trilayers of sputtered $Al_{2}O_{3}/Pt/Au $ of respective thicknesses 5, 3 and 200 nm. This procedure insures that the tubes do not contain any chemical dopants such as alkalis or halogens. The contacts showed no sign of superconductivity down to 50 mK. The samples were measured in a dilution %\linebreak
\begin{figure}[htb]
\vspace{-3cm}
\begin{center}
\includegraphics[clip=true,width=8cm]{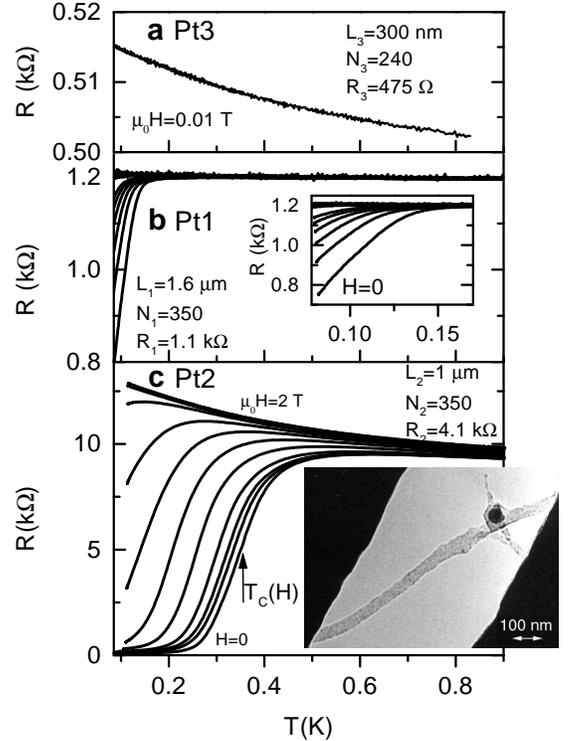}
%\leavevmode
%\epsfxsize=8cm
%\epsfbox{fig1.eps}
\end{center}
%\vspace{0 cm}
\caption{
\label{figure1}
Resistance as a function of temperature for the three samples. 
The length L, number of tubes N and room temperature resistance R of each sample are given in the corresponding panel. a: Sample Pt3. b: Resistance of Pt1 in applied magnetic fields of $\mu_{0}H$= 0, 0.02, 0.04, 0.06, 0.08, 0.1, 0.2, 0.4, 0.6, 0.8 and 1 T from bottom to top. Inset is a zoom of the low temperature region. c: Resistance of Pt2 at $\mu_{0}H$=0, 0.05, 0.1, 0.2, 0.4, 0.6, 0.8, 1, 1.25, 1.5, 1.75, 2, 2.5 T from bottom to top. 
Inset: TEM micrograph of sample Pt2, from which we deduce $L_{2}$ and $N_{2}$. $N_{2}$ is estimated from the measured diameter $D_{2}$, through $N_{2} =(D_{2} /(d+e))^2$, where d is the diameter of a single tube (d=1.4 nm), and e is the typical distance between tubes in a rope (e=0.2 nm). The dark spot is a Ni/Y catalyst particle.} 
\end{figure}

\noindent
refrigerator, at temperatures ranging from 1~K to 0.05~K, through filtered lines \cite{Reulet2}. Magnetic fields up to 5 T could be applied perpendicularly to the contacts and the tubes. The resistance was measured by applying a small (1 nA to 10 nA, 30 Hz) a.c. current though the sample and measuring the a.c. voltage using lock-in detection.

We select samples with a room temperature (RT) resistance less than 10 k$\Omega$. As is generally observed, we find that the resistance increases as the temperature is lowered between 300 K and 1K . Things change however below 1K, as shown in Fig. 1 for the three samples Pt1, Pt2, and Pt3, measured in magnetic fields ranging from 0 to 2.5 T. At zero field, the zero-bias resistance of Pt3 increases as T is reduced, whereas the resistances of Pt1 and Pt2 decrease drastically below $T_{1}^*= 140$ mK for Pt1 and $T_{2}^*= 550$ mK for Pt2. The resistance of Pt1 is reduced by 30\% at 70 mK. That of Pt2 decreases by more than two orders of magnitude, and saturates below 100 mK at a  value $R_{r}=74$~$\Omega$. We define a transition temperature $T_{C_{2}}$ by the inflexion point of R(T). $T_{C_{2}}$  is 370 mK at zero field,  decreases at higher magnetic fields, and extrapolates to zero at 1.35 T (Fig 4c). At fields above 1.25 T, the  resistance increases with decreasing temperature, similarly to Pt3, and becomes independent of magnetic field. The  resistance of Pt1 follows qualitatively the same trend, but the full transition did not occur down to 70 mK. Figures 2  and 3 show that in the temperature and field range where the zero-bias resistance drops, the differential resistance is  strongly bias-dependent, with lower resistance at low bias.
	These data suggest that the rope Pt2 (and, to a lesser extent, Pt1) is intrinsically superconducting. Although the experimental data of Pt2 seem similar to those of SWNT connected to superconducting contacts \cite{Kasumov},  there are major differences. In particular the $V(I), dV/dI(I)$ do not show any supercurrent because of the existence of a finite residual resistance. 

We now analyse the superconductivity in these systems, taking into account several features: the large normal contacts, the coupling between tubes within the rope, the 1D character of each tube, and their finite length compared to relevant mesoscopic and superconducting scales.
The resistance of any superconducting wire measured through normal contacts (an NSN junction) cannot be zero because the number of channels in the wire is much smaller than in the contacts\cite{Landauer}: a metallic SWNT, with 2 conducting channels, has a contact resistance of half the resistance quantum, $R_{Q}/2$ (where $R_{Q}=h/(2e^2)$=12.9 k$\Omega$), even if it is superconducting. A rope of  $N_{m}$ parallel metallic SWNT will have a minimum resistance of $R_{Q}/(2N)$. Therefore we use the residual resistance $R_{r}=74$~$\Omega$ of Pt2 to deduce that Pt2 has at least $N_{m}=R_{Q}/2R_{r}\approx 90$ metallic tubes. This is approximately one quarter of the number of tubes in the rope, measured by TEM (Fig 1c).
Similarly, $R_{Q}$  is also the maximum resistance of any phase coherent metallic wire\cite{Thouless}. As a consequence, the high value (9.2 k$\Omega$) of the resistance at 1K (which corresponds to an average resistance per metallic tube of $9.2$ k$\Omega*N_{m}=830$~k$\Omega=130$~$R_{Q}$) cannot be understood if the nanotubes are independent, unless considering a very short (unphysical) phase coherence length $L_{\varphi}(1K)=L/130=$ 8 nm. On the other hand if the electrons are free to move from tube to tube\cite{Maarouf}, the resistance is simply explained by the presence of disorder. The mean free path is deduced from the RT resistance $R_{2}=4.1$ k$\Omega$   through \cite{Imry} $l_{e2}\approx \frac{L}{R_{2}}\frac{R_{Q}}{N_{m}}\approx 18$ nm. We conclude that Pt2 is a diffusive conductor with a few hundred conduction channels.

With such a small number of channels, we expect the superconductivity to differ from 3D superconductivity. In particular, we expect to observe a broad resistance drop starting at the 3D transition temperature \cite{Giordano} $T^* $ and going eventually to $R_{r}$ at zero temperature. This is what is observed in Pt2 (see figure 1.c). We estimate the gap through the BCS relation $\Delta =1.76$ $k_{B}T^*$ : $\Delta \approx 85$~$\mu eV$ for Pt2. We can then deduce the superconducting coherence length along the tube in the diffusive limit $\xi_{2}=\sqrt{\hbar v_{f}l_{e}/\Delta}\approx 0.3$~$\mu$m where $v_{f}$ is the longitudinal Fermi velocity $8\times10^5$ m/s \cite{Bourbonnais}. Consistent with 1D superconductivity, $\xi_{2}$ is ten times larger than the diameter of the rope. 

\begin{figure}[hbt]
\begin{center}
%\vspace{1 cm}
%\leavevmode
%\epsfxsize=7.0cm
%\epsfbox{fig2.eps}
\includegraphics[clip=true,width=7cm]{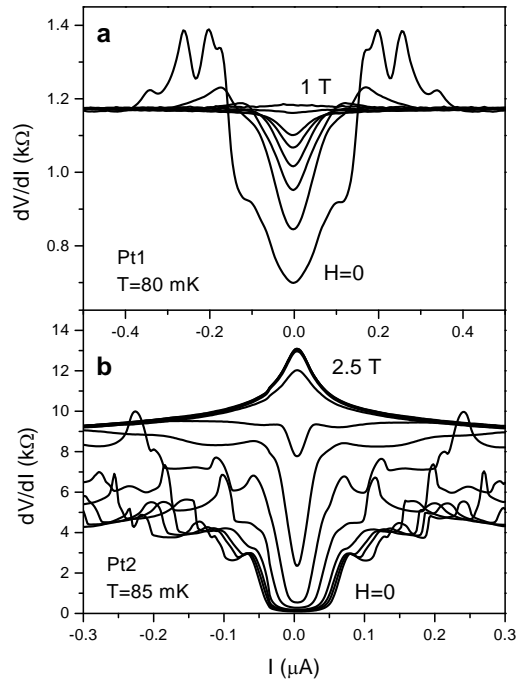}
\end{center}
\caption{
\label{figure2}
Differential resistance as a function of current for samples Pt1 and Pt2, in different applied fields. a: Sample Pt1. Fields are 0, 0.02, 0.04, 0.06, 0.08, 0.1, 0.2 and 1 T. b: Sample Pt2. Fields are 0, 0.2, 0.4, 0.6, 0.8, 1, 1.25, 1.5, 1.75, 2, and 2.5 T.}
\end{figure}

%\linebreak
\begin{figure}[htb]
\begin{center}
%\leavevmode
%\epsfxsize=8 cm
%\epsfbox{fig3.eps}
%\vspace{0 cm}
\includegraphics[clip=true,width=8cm]{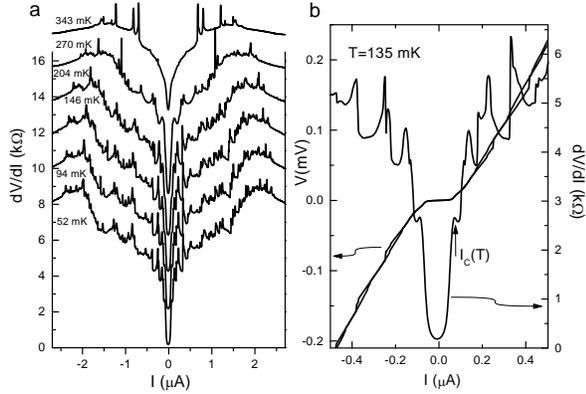}
\end{center}
\caption{
\label{figure3}
Left panel: Differential resistance of Pt2 vs. current for a larger current amplitude than in Figure 2, at different temperatures. Curves are offset vertically for clarity. Right panel: V(I) and $\frac{dV}{dI}(I)$ curves showing the hysteretic behavior in V(I) at each peak in the $\frac{dV}{dI}(I)$  curve.}
\end{figure}

Finally, reminiscent of measurements of narrow superconducting metal wires \cite{Giordano}, we find jumps in the differential resistance as the current is increased (Figures 2 and 3). For Pt2 the differential resistance at low currents remains equal to $R_{r}$ up to 50 nA, where it strongly rises but does not recover its normal state value until 2.5 $\mu$A (fig 3a). 
The jump in resistance at the first step corresponds approximately to the normal state resistance of a length $\xi_{2}$ of Pt2. Each peak corresponds to a hysteretic feature in the V-I curve (fig 3b). Above 1 T the differential resistance is peaked at zero current. This is also the case for Pt3 (data not shown). The variations of the differential resistance of Pt1 are similar to those of Pt2 close to its transition temperature. These  jumps are identified as phase slips \cite{Giordano,Meyer,Tinkham}, which are the occurrence of normal regions located around defects in the sample. Such phase slips can be thermally activated (TAPS), leading to an exponential decrease of the resistance instead of a sharp transition, in qualitative agreement with our experimental observation (fig 4a). At sufficiently low temperature, TAPS are replaced by quantum phase slips (QPS), which, when tunneling through the sample, contribute an additional resistance to the zero temperature resistance. Moreover, QPS are predicted to supress the transition when the normal state resistance of the sample on the phase coherence scale is larger than $R_{Q}/2$  \cite{Zaikin}(as confirmed by recent experiments \cite{Bezryadin}). Our data on Pt2 show no evidence of such an effect, even though the normal state resistance, measured above T*, is 40\% larger than $R_{Q}/2 $.
% A possible explanation is that $L_{\varphi}$ is slightly smaller than $L_{2}$ in this relatively long sample. 
	The current above which the jumps disappear, 2.5 $\mu A$, is close to the critical current $I_{C}=\Delta /R_{r}e \approx 1$ $ \mu$A of a superconducting wire without disorder and with the same number of conducting channels \cite{Tinkham}. This large value of critical current would also be the maximum supercurrent in a structure with this same wire placed between superconducting contacts (with gap $\Delta_{S}$), and is much larger than the Ambegaokar-Baratoff  %\linebreak
\begin{figure}
\begin{center}
%\leavevmode
%\epsfxsize=6 cm
%\epsfbox{fig4.eps}
\vspace{.8 cm}
\includegraphics[width=7cm]{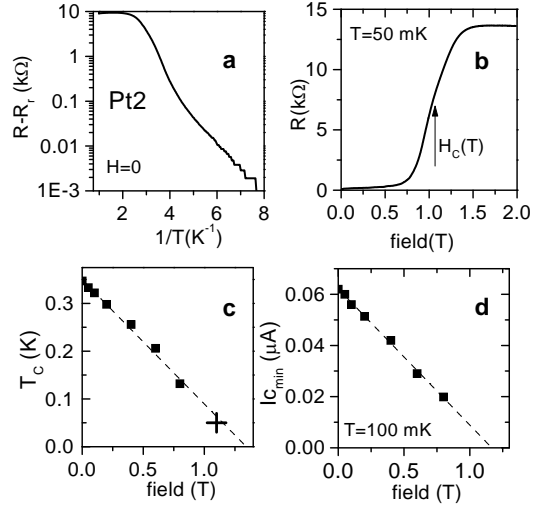}
\end{center}
\caption{
\label{figure4}
a Resistance of Pt2 plotted on a log scale as a function of the inverse temperature at H=0. We have subtracted the low temperature residual resistance (contact resistance). The slope yields an approximate activation energy of 0.8 K. 
b Magnetoresistance of Pt2 at 50 mK. We define the critical field as the inflection point of R(H):  $\mu_{0}H_{C}$(T= 50 mK)=1.1 T. c Transition line of Pt2 defined in the H,T plane by the inflection point of R(T) or equivalently by the inflection point of R(H). d Field dependence of the critical current of Pt2 defined as the current at which the first resistance jump occurs in the dV/dI curves of Fig. 2. $I_{C}(H)$ extrapolates to a critical field of 1.2 T, in agreement with the linear extrapolation 1.3 T of $T_{C}(H)$.
}
\end{figure}   

\noindent   
prediction $R_{N} I_{C}= \Delta_{S}/e$. This might explain the anomalously large supercurrent measured in a previous experiment \cite{Kasumov}, where nanotubes were connected to superconducting contacts.

We now discuss the effect of the magnetic field. The field at which the resistance saturates to its normal value and at which the critical current vanishes, 1.25 T, coincides with the field obtained by extrapolation of $T_{C}(H)$ to zero temperature (fig. 4b). 
It is difficult to say what causes the disappearance of superconductivity. The value of Hc(0) should be compared to the depairing field in a confined geometry \cite{Meservey}, and corresponds to a flux quantum $\Phi_{0}$ through a length $\xi$ of an individual SWNT of diameter d, $\mu_{0}H_{C}=  \Phi_{0}/(2\sqrt{\pi}d\xi)=1.35$ T. But $H_{C}(0) $is also close to the field $\mu_{0}H_{p}=\Delta /\mu_{B}=1.43$ T at which a paramagnetic state becomes more favorable than the superconducting state \cite{Clogston,Chandrasekhar}. Note that this value is of the same order as the critical field that was measured on SWNT connected between superconducting contacts, i.e. much higher than the critical field of the contacts. 

We now estimate the superconducting coherence length of the two other samples, to explain the extent or absence of observed transition. Indeed, investigation of the proximity effect at high-transparency NS interfaces has shown that superconductivity resists the presence of normal contacts only if the length of the superconductor is much greater than $\xi$ \cite{Belzig}, i.e. if the wire contains a superconducting reservoir. This condition is nearly fulfilled in Pt2 ($\xi_{2}  \approx L_{2}/3$). Using the high temperature resistance values of Pt1 and Pt3, and assuming a gap $\Delta$ and number of metallic tubes equal to those of Pt2 we find $\xi_{1} \approx L_{1}/2$ and $\xi_{3}  \approx 2L_{3}$. These values explain qualitatively a reduced transition temperature for Pt1 and the absence of a transition for Pt3. Moreover we can argue that the superconducting transitions we see are not due to a hidden proximity effect : if the $Al_{2}O_{3}/Pt/Au$ contacts were made superconducting by the laser pulse, the shortest nanotube (Pt3) would become superconducting at temperatures higher than the longer tubes (Pt1 and Pt2). The main result, i.e. no visible transition with a short rope, and a visible transition in a long rope, are confirmed by measurements on two other samples which are not presented here.

We now consider the possible mechanism of superconductivity. It has been suggested that coupling with low energy phonons can turn repulsive interactions in a Luttinger liquid into attractive ones and drive the system towards a superconducting phase \cite{Loss}. Such low energy phonons have been experimentally observed in the form of mechanical bending modes of a suspended SWNT rope \cite{Reulet}. It was also shown that superconducting fluctuations can dominate at low temperature in ladders such as tubes \cite{Egger}. In this case the system must be away from half-filling, a condition probably fulfilled in our experimental situation, due to hole doping from the contacts \cite{Venema,Odinstov}. Finally, the superconductivity reported here recalls that of graphite intercalated with alkalis (Cs,K), which also occurs between 0.2 and 0.5 K \cite{Hannay}. Much higher temperatures were observed in alkali doped fullerenes \cite{Gunnarsson} because of the coupling to higher energy phonons. This suggests the possibility of increasing the transition temperature by chemically doping the nanotubes.

We have shown that ropes of carbon nanotubes are intrinsically superconducting. This is the first observation of superconductivity in a system with such a small number of conduction channels. The understanding of this superconductivity calls for future experimental and theoretical work and motivates in particular a search of superconducting fluctuations in a single SWNT.

A.K. thanks the Russian foundation for basic research and solid state nanostructures for financial support, and thanks CNRS for a visitor's position. We thank M. Devoret, N. Dupuis, T. Martin, D. Maslov, C. Pasquier for stimulating discussions.

\end{multicols}

\vspace{-.5cm}
\begin{thebibliography}{99}
\bibitem{Dresselhaus}M.S. Dresselhaus, G. Dresselhaus, and P.C. Eklund, {\it Science of Fullerenes and Carbon nanotubes} (Academic, San Diego, 1996).
\bibitem{Hamada} N. Hamada, S. I. Sawada and A. Oshiyama, Phys.  Rev.  Lett. {\bf 68}, 1579 (1992).
\bibitem{Wildoer} J. W. G. Wildoer {\it et al.}, Nature {\bf 391}, 59 (1998).
 \bibitem{Egger} R. Egger, A. Gogolin,  Phys. Rev. Lett. {\bf 79}, 5082 (1997). R. Egger, Phys. Rev. Lett. {\bf 83}, 5547 (1999).
 \bibitem{Kane} C. Kane, L. Balents, M. P. Fisher, Phys. Rev. Lett {\bf 79}, 5086 (1997).
 \bibitem{Bockrath} M. Bockrath {\it et al.}, Nature {\bf 397}, 598 (1999).
 \bibitem{Grabert} H. Grabert and M. H. Devoret (eds), Single Charge Tunneling (Plenum, New-York, 1992); J. T. Tans {\it et al.}, Nature {\bf 386}, 474 (1997).
  \bibitem{Kasumov2} A.Yu. Kasumov, I.I. Khodos, P.M. Ajayan, C. Colliex, Europhys. Lett. {\bf 34}, 429 (1996); A.Yu. Kasumov {\it et al.}, Europhys. Lett. {\bf 43}, 89 (1998).
\bibitem{Kasumov} A.Yu. Kasumov {\it et al.}, Science {\bf 284}, 1508 (1999).
\bibitem{Journet} C. Journet,  {\it et al.}, Nature {\bf 388}, 756 (1997).
 \bibitem{Vaccarini} L. Vaccarini, {\it et al.}, C.R.Acad.Sci.{\bf 327}, 925 (1999).
 \bibitem{Reulet2} B. Reulet, H. Bouchiat, and D. Mailly, Europhys. Lett. {\bf 31}, 305 (1995).
 \bibitem{Landauer} R. Landauer, IBM Res. Dev. 1, 223 (1957). 
 \bibitem{Thouless} D. J. Thouless, Phys. Rev. Lett. {\bf 39}, 1967 (1977).
 \bibitem{Maarouf} Electronic transfer between tubes in a disordered rope is indeed suggested in: A. A. Maarouf, C. L. Kane, and E. J. Mele,  Phys. Rev. B{\bf 61}, 11156 (2000);  H.R. Shea, R. Martel and Ph. Avouris, Phys. Rev. Lett. {\bf 84}, 4441 (2000).
  \bibitem{Imry} Y. Imry, Europhys. Lett. {\bf 1}, 249 (1986).
 \bibitem{Giordano} N. Giordano, Phys. Rev. B {\bf 50}, 160 (1991).
\bibitem{Bourbonnais} Note however that the transverse velocity is certainly much smaller, yielding a transverse coherence length $\xi_{\perp}$ smaller than the longitudinal one, but the 1D character of the transition depicted below indicates that  $\xi_{\perp}$ is certainly larger than the diameter of the rope. Such anisotropic transport is observed in organic conductors where a very small coupling between 1D chains restores a 3D behavior of the electrons, and allows superconductivity in a macroscopic sample. (see e.g. C. Bourbonnais,  D. J\'er\^ome, in Advance in Synthetic Metals, eds P.Bernier, S. Lefrant and G. Bidan) 206 (1998).
 \bibitem{Meyer} J. Meyer, G. V. Minnigerode, Physics Letters, {\bf 38A}, 7, 529 (1972).
 \bibitem{Tinkham} Tinkham, M., Introduction to superconductivity, McGraw-Hill, 2d Ed. (Singapore, 1996).
  \bibitem{Zaikin} A.D. Zaikin {\it et al.}, Phys. Rev. Lett. {\bf 78}, 1552 (1997).
 \bibitem{Bezryadin} A. Bezryadin, C. N. Lau and M. Tinkham Nature {\bf 404}, 971 (2000).
 
 \bibitem{Meservey} R. Meservey, P.M. Tedrow, Phys. Rep. {\bf 238}, no.4, 173 (1994).

 \bibitem{Clogston} A.M. Clogston, Phys. Rev. Lett. {\bf 9}, 266 (1962).

 \bibitem{Chandrasekhar} B.S. Chandrasekhar, Appl. Phys. Lett. {\bf 1}, 7 (1962).

 \bibitem{Belzig} W. Belzig, C. Bruder and G. Sch$\ddot{o}$n, Phys. Rev. B {\bf 54}, 9443 (1996).
  \bibitem{Loss} D. Loss, and T. Martin, Phys. Rev. B {\bf 50}, 12160 (1994).

 \bibitem{Reulet} B. Reulet {\it et al.}, Phys. Rev. Lett {\bf 85},  2829 (2000).

 \bibitem{Venema} L. C. Venema {\it et al.}, Science {\bf 283}, 52 (1999).

 \bibitem{Odinstov} A.A. Odintsov , Phys. Rev. Lett. {\bf 85}, 150 (2000).
 \bibitem{Hannay} N. B. Hannay {\it et al.}, Phys. Rev. Lett. {\bf 14}, 225 (1965).
 \bibitem{Gunnarsson} O. Gunnarsson, Rev. Mod. Phys. {\bf 69}, 575 (1997).
 \end{thebibliography}
\end{document}